%Paper: hep-th/9410123
%From: stephany@usb.ve (Jorge Stephany Ruiz)
%Date: Tue, 18 Oct 1994 11:36:45 -0400 (GMT-0400)

\input amstex
\magnification=1200
\TagsOnRight
\def\wh{\widehat}

\def\D{\Cal{D}}

\def\ov{\overline}

\def\noi{\noindent}

\mathsurround=1pt
\tolerance=10000
\pretolerance=10000
\nopagenumbers
\headline={\ifnum\pageno=1\hfil\else\hss\tenrm -- \folio\ --\hss\fi}
\hsize=16 true cm
\vsize=23 true cm

\line{Preprint {\bf SB/F/94-225}}
\hrule
\vglue 1.5cm
\vskip 2.cm

\centerline{\bf A NEW ACTION PRINCIPLE FOR WITTEN'S }
\centerline{\bf TOPOLOGICAL FIELD THEORY  }

\vskip 1.5cm
\centerline{ R. Gianvittorio, A. Restuccia and J. Stephany}
\vskip .5cm
\centerline{\it Universidad Sim\'{o}n Bol\'{\i}var, Departamento de
F\'{\i}sica}
\centerline{\it Apartado postal 89000, Caracas 1080-A, Venezuela.}
\centerline{\it e-mail: ritagian{\@}usb.ve, arestu{\@}usb.ve,
stephany{\@}usb.ve}
\vskip 1cm
{\bf Abstract}
\vskip .5cm
{\narrower{\flushpar In this letter a new gauge invariant, metric
independent action is introduced from which  Witten's Topological Quantum
Field Theory may be obtained after gauge fixing using standard BRST
techniques. In our model the BRST algebra of transformations, under which
the effective action is invariant, close off-shell in distintion with what
occurs in the one proposed by Labastida and Pernici. Our approach provides
the geometrical principle for the quantum theory. We also compare our
results with an alternative formulation presented by  Baulieu and
Singer.\par}}   \vskip 3cm

\hrule
\bigskip
\centerline {\bf UNIVERSIDAD SIMON BOLIVAR}

\newpage

The physical and mathematical properties of topological field
theories have   raised  much interest in recent times (see for example
 Refs.[1-5]). In this letter we discuss an alternative approach
for the construction of Witten's Topological Quantum Field Theory (TQFT)
[1]  using BRST techniques and starting from a new action principle
different from the ones discussed in Refs.[2] and [3].

   In Ref.[1] starting from  a BRST-like supersymmetric principle Witten
constructed a quantum effective action  from which Donaldson
invariants may be computed as correlation functions of adequatly choosen
observables. In particular he showed, in the context of quantum field theory,
the relation of Floer homology and Donaldson invariants.

Specifically the topological invariant path integral formula he obtained was

$$
Z(\gamma_1 ,\cdots,\gamma_N )=\int_{M} \D x \ exp({-L'\over{e^2}})
\prod^N_{i=1} \int_{\gamma_i} W_{k_i} ,\tag 1
$$
where $\gamma_i$ is a $k_i$ dimensional homology cycle on $M$ and
$W_{k_i}$ are structure functionals obtained from
$W_0(P)={1\over{2}}Tr{\phi}^2(P)$, $\phi$ being the spin zero field in
his gauge fixed action. In (1) $L'$ is the gauge fixed Lagrangean obtained from
SUSY arguments ( see eq. (2.41) in [1]).

In the same paper [1] the problem of finding a manifestly generally covariant
gauge theory whose BRST gauge fixing gives this gauge fixed Lagrangean was
raised.
Since then much work has been done to resolve this issue and important steps
in the correct direction have been performed. The discussion on this point
has been concentrated around two different actions proposed in Ref.[2] and
Ref.[3] which have been shown to be related through the BRST formalism
with Witten's TQFT.

The gauge invariant action proposed by Labastida and Pernici in [2] does indeed
give rise after BRST gauge fixing to a TQFT which is identical to Witten's
theory. The effective Lagrangean in [2], (see eq. (14) in [2]), is almost
identical to $L'$. It becomes identical when a sligth change in the gauge
fixing
conditions is performed (we discuss that point later on). Nevertheless
since it is not independent of the metric on $M$, it do not represents
the truly geometrical principle for the theory and the BRST algebra
obtained in [2] by following the BV approach only closes on-shell.

In [3] the same problem  was addressed with emphasis in
exhibit a manifestly covariant gauge action, independent of the metric on $M$.
The action proposed by Baulieu and Singer in [3], was the Pontryagin invariant
$$ {1\over{4}}\int_{M} F\wedge F.\tag 2 $$

Its BRST gauge fixing leads to an effective Lagrangean $L$, (see eq.(2.13)
in [1]) from which $L'$ is obtained only after addition of a BRST invariant
term of the same form as (2), that is
$$
L' = L + \alpha F\wedge F .
$$

This last step which is completely valid in order to
study the Donaldson invariants from a quantum field theory point of view,
is an ad hoc modification of the BRST procedure and in fact does not
resolve the problem we are considering.

In what follows we introduce a new manifestly covariant gauge action,
{\it independent of the metric on M, whose BRST gauge fixing leads
exactly to Witten's TQFT}.  The classical field equations we obtain are
the same as in [2] but with the important difference mentioned above that they
now arise from a variational principle of a metric independent gauge invariant
action. Moreover the canonical formulation of the new action we are proposing
 also agrees exactly with the canonical formulation of the action in [2].
The effective lagrangean that we obtain is identical to Witten«s $L'$.

The action we propose is given by
$$
S={1\over{4}}\int Tr(B+F)\wedge (B+F) ={1\over{4}}\int
d^4x \
%% FOLLOWING LINE CANNOT BE BROKEN BEFORE 80 CHAR
\epsilon^{\mu\nu\sigma\rho}(F_{\mu\nu}^a+B_{\mu\nu}^a)(F_{\sigma\rho}^a+B_{\sigma\rho}^a).\tag3
$$   where $F_{\mu\nu}^a$ is the curvature of the gauge connection
$A_{\mu}^a$, with space-time index $\mu$ and group index $a$;
$B_{\mu\nu}^a$ is an independent antisymmetric field. It is important to
mention that Horowitz in Ref.[7] introduces another topological gauge
invariant action in terms of the same fields  $B$ and $F(A)$, which
nevertheless is BRST equivalent to take the Pontryagin invariant as was
done by Baulieu and Singer in Ref.[3].

The action (3) is invariant under the infinitesimal non-abelian gauge
transformations
$$
\align
&\delta A_\mu^a =\epsilon_\mu^a ,\\
&\delta B_{\mu\nu}^a = -\D_{[\mu}\epsilon_{\nu ]}^a .\tag 4
\endalign
$$
with $\epsilon_\mu^a$ the gauge parameters.

The field equations associated to (3) are
$$
F_{\mu\nu}^a+B_{\mu\nu}^a = 0 .\tag 5
$$

Evaluated over the classical solutions of the field equations, our action
and the one proposed in Ref.[2] are zero . This is not the
case before quantization for the action proposed in (2). In this case
the solutions to the field equations are either  selfdual or antiselfdual
and evaluation of the action gives the corresponding Pontryagin index.

The gauge invariances of the action (3) allow to make the partial gauge
fixation  $$ B_{\mu\nu}^{a+}=0 , \tag 6a
$$
where
$$
B_{\mu\nu}^{a(\pm)} \equiv {1\over{2}} (B_{\mu\nu}^a \pm \ov{B}_{\mu\nu}^a)
, \tag 6b $$
$$
\ov{B}_{\mu\nu}^a \equiv {1\over{2}}
\epsilon_{\mu\nu\sigma\rho}B^{\sigma\rho a}. \tag 6c
$$

The field equations then reduce to
$$
F_{\mu\nu}^{a+}=0, \tag 7a
$$
$$
F_{\mu\nu}^{a-}+B_{\mu\nu}^{a-}=0, \tag 7b
$$
which are invariant under the remaining gauge freedom of (3), that is
$$
\align
&\delta A_\mu^a = \D_\mu\epsilon^a ,\\
&\delta B_{\mu\nu}^a = 0. \tag 8
\endalign
$$

This field equations (7) are identical to the ones obtained in [2] by Labastida
and Pernici. They have been obtained in our case from an action which is
manifestly independent of the metric.

The action (3) may be also formulated in a canonical form. Then we obtain
$$
S=\int
d^4x \
%% FOLLOWING LINE CANNOT BE BROKEN BEFORE 80 CHAR
\dot{A}_i^a\epsilon^{ijk}(F_{jk}^a+B_{jk}^a)+A_0^a\D_i[\epsilon^{ijk}(F_{jk}^a+B_{jk}^a)]+B_{0i}^a
\epsilon^{ijk}(F_{jk}^a+B_{jk}^a). \tag 9  $$
Here  we recognize a theory
with vanishing canonical Hamiltonian and with canonical conjugate momenta
to $A_i^a$ given by $$
\pi^{ia}=\epsilon^{ijk}(F_{jk}^a+B_{jk}^a), \tag 10
$$
where $\epsilon^{ijk}\equiv \epsilon^{0ijk}$, and with Lagrange multipliers
$A_0^a$ and $B_{0i}^a$ associated respectively to the following constraints
$$\align
&\phi^a =
(\D_i\pi^i)^a=\partial_i\pi^{ia}+(A_i\times
\pi^i)^a=\partial_i\pi^{ia}+f^{abc}A_i^b\pi^{ic}=0,\tag 11a\\  &\phi^{ia}
=\pi^{ia}=0.\tag 11b \endalign $$

The algebra of the constraints is given by:
$$\align
&\{\phi^{ia}(x),\phi^{jb}(x')\} =0,\\
&\{\phi^a(x),\phi^b(x')\} =f^{abc}\phi^c(x)\delta^3(x-x'),\\
&\{\phi^a(x),\phi^{ib}(x')\} =f^{abc}\Phi^{ic}(x)\delta^3(x-x'),\tag 12
\endalign
$$
which show that all constraints are first class. Nevertheless they are not
linearly independent since they satisfy the following identity
$$
(\D_i\phi^i)^a=\phi^a,\tag 13
$$
and thus we are in front of a reducible theory [8,9] with one
level of reducibility. The corresponding the matrix of reducibility is given
by:  $$ a=(\D_i,-1).\tag 14 $$

As we already mentioned one can show after several calculations that the
canonical formulation of the action proposed in [2] exactly agrees with the
results discussed above. It is interesting to note that the explicit
dependence on the metric of the action of Labastida and Pernici [2]
disappears in the canonical formulation, explaining from a different point
of view why this theory is  topological. The dependence on the metric of
the associated functional integral may only arise from the gauge fixing
functions but of course the functional integral is itself independent of
the gauge fixing conditions.

It is also interesting to remark that the algebra of first class constraints
(12) is the same for the theories proposed in Ref. [2] and
Ref.[3] and also for the 4-dim BF topological field theory Ref.[10] whose
action is given by
$$
\int_{M} B\wedge F . \tag 15
$$
This action in principle is not related to Witten's TQFT.
In facts the classical solutions of the BF theory are flat connections
belonging to the different theta sectors. This space of
classical solutions is thus different from the one associated to Witten's
theory. The BF theory is usually classified [5] as a type Schwarz [4]
topological theory in distinction to the Witten's type topological theories
[1]. This distinction however does not seem to be relevant in this case.
All the properties of both theories arises from the general BRST treatment
of the corresponding {\it metric independent} gauge invariant actions.
This include the property that the semiclassical aproximation is exact as
we will discuss below.

To construct the BRST charge we follow Ref.[9] and introduce the minimal
sector of the extended phase space expanded by the conjugate pairs:
$$
%% FOLLOWING LINE CANNOT BE BROKEN BEFORE 80 CHAR
(A_i^a,\pi^{ia});(C_1^a,\mu^{1a}),(C_{1i}^a,\mu^{1ia});(C_{11}^a,\mu^{11a}),\tag 16
$$
where $(A_i^a,\pi^{ia})$ are the original canonical coordinates and
$(C^a,\mu^a)$
are the canonical ghost and antighost associated to the constraints (11).

The off-shell nilpotent BRST charge is then given by:
$$
\align
\Omega=<&
C_1^a(\D_i\pi^i)^a+C_{1i}^a\pi^{ia}+C_{11}^a[(\D_i\mu^{1i})^a-\mu^{1a}]\\
&-{1\over{2}}C_1^aC_1^bf^{abc}\mu^{1c}-C_1^aC_{1i}^bf^{abc}\mu^{1ic}\\
&-C_{11}^aC_1^bf^{abc}\mu^{11c}>,\tag 17
\endalign
$$
where $<\cdots >$ stands for integration on the space like continuous
index.
As we said, since the first class constraints of [2] are identical to (11)
the BRST charge is the same. The Hamiltonian is zero, as it
should be for a topological field theory.

We now define the non minimal sector of the
extended phase space [9]. It contains extra ghosts, antighosts and Lagrange
multipliers. First we introduce the C-fields
$$
 C_m^a,C_{mi}^a;\ \ \
C_{mn}^a,C_{mni}^a;\ \ \ \ \ \ m,n=1,2,3\tag 18
$$
where at least one of the indices $m,n$ take the values 2 or  3. In
addition to these ghost, antighost and  Lagrange multiplier fields we
introduce the $\lambda$ and $\theta$ fields  (Lagrange multipliers),
also in the non minimal sector, $$
\align
&\lambda_1^{0a},\ \lambda_{1i}^{0a};\ \lambda_{1m}^{0a};\ \ m=1,2,3\\
&\lambda_{11}^{1a};\\
&\theta_1^{0a},\ \theta_{1i}^{0a};\ \theta_{1m}^{0a};\ \ m=1,2,3\\
&\theta_{11}^{1a}. \tag 19
\endalign
$$

In this notation the 1 subscripts denote ghost associated to a gauge
symmetry of the action, the 2 subscripts denote antighost associated to a gauge
fixing  condition in the effective action and the 3 subscripts denote
Lagrange  multipliers associated to a gauge fixing condition.
The effective action is then given by:
$$
\align
%% FOLLOWING LINE CANNOT BE BROKEN BEFORE 80 CHAR
Seff=\int^{tf}_{ti}dt[&\pi^{ia}\dot{A}_i^a+\mu^{1a}\dot{C}_1^a+\mu^{1ia}\dot{C}_{1i}^a+\mu^{11a}\dot{C}_{11}^a
%% FOLLOWING LINE CANNOT BE BROKEN BEFORE 80 CHAR
+\\&\wh{\delta}(\lambda_1^{0a}\mu^{1a}+\lambda_{1i}^{0a}\mu^{1ia}+\lambda_{11}^{1a}\mu^{11a})+L_{GF+FP}],\tag
20
\endalign
$$
where
$$
%% FOLLOWING LINE CANNOT BE BROKEN BEFORE 80 CHAR
L_{GF+FP}=\wh{\delta}(C_2^a\chi_2^a+C_{2i}^a\chi_2^{ia})+\wh{\delta}(\sum^3_{m=1}C_{m2}^a\chi_{m2}^a)+
\wh{\delta}(\lambda_{12}^{0a}\Lambda_2^a+\theta_{12}^{0a}\Theta_2^a),\tag
21
$$
is the sum of the generalizations of the Fadeev-Popov and gauge fixing
terms. In Eq.(21) $\chi_2^a$, $\chi_2^{ia}$ are the primary gauge fixing
functions associated  to the constraints (11), while $\chi_{m2}^a$,
$\Lambda_2^a$ and $\Theta_2^a$ are gauge fixing functions which must fix the
longitudinal part of fields of the non minimal sector. The BRST
transformation for the canonical variables is given by
$$
\wh{\delta}Z=(-1)^{\epsilon_z}\{ Z,\Omega \},\tag 22
$$
where $\epsilon_z$ is the grassmanian parity of $Z$. The BRST
transformation of the variables of the non minimal sector are fixed
imposing the closure of the charge as in Ref.[9]. After some
simplifications we finally choose gauge fixing functions that may be
written in a covariant form as $$
\align
&\chi_2^a=\partial_{\mu}A^{\mu a}-{\alpha\over{2}}C_3^a,\\
&\chi_2^{\mu\nu a}={1\over{2}}
\epsilon^{\mu\nu\sigma\rho}B_{\sigma\rho}^a+B^{\mu\nu a},\\
&\chi_{12}^a=\D^{\mu}C_{1\mu}^a+{1\over{2}}
C_{13}^a \times C_{11}^a+{1\over{2}}((C_{12}\times C_1)\times C_{11})^a, \tag
23
\endalign
$$
where $C_{1\mu}^a=(-\lambda_{11}^{1a},C_{1i}^a)$.
After elimination of all conjugate momenta in the functional integral, the
BRST transformation rules of all the remaining geometrical objects are
covariant and take the form
$$\align
&\wh{\delta}A_{\mu}^a=-\D_{\mu}C_1^a+C_{1\mu}^a,\\
&\wh{\delta}C_1^a=C_{11}^a+{1\over{2}}(C_1\times C_1)^a,\\
&\wh{\delta}C_{1\mu}^a=\D_{\mu}C_{11}^a+(C_1\times C_{1\mu})^a,\\
&\wh{\delta}C_{11}^a=-(C_{11}\times C_1)^a,\\
&\wh{\delta}C_2^a=C_3^a,	\	\ \ \ \	\	\	\	\ \wh{\delta}C_3=0,\\
&\wh{\delta}C_{2\mu\nu}^a=C_{3\mu\nu}^a,	\	\	\	\	\	\
\wh{\delta}C_{3\mu\nu}=0,\\
&\wh{\delta}C_{12}^a=C_{13}^a,	\	\ \ \	\	\	\	\ \wh{\delta}C_{13}=0. \tag 24
\endalign
$$

The BRST invariant action, once we have eliminated $B_{\mu\nu}^a$, may be
written as
$$
S=S_0+S_1+S_2, \tag 25
$$
where
$$
S_0=<{1\over{4}}F_{\mu\nu}^{a+}F^{\mu\nu a+}>,\tag 26
$$
$$\align
S_1=<&-C_2^{\mu\nu
a}\D_{\mu}C_{1\nu}^a+{1\over{8}}C_{11}^a(C_2^{\mu\nu}\times
C_{2\mu\nu})^a+\ov{C}_{13}^a\D_{\mu}C_1^{\mu
a}\\ &+C_{12}^a(C_1^\mu \times C_{1\mu})^a+C_{12}^a\D_{\mu}\D^\mu C_{11}^a\\
&+{1\over{2}}C_{11}^a(\ov{C}_{13}\times \ov{C}_{13})^a-{1\over{2}}(C_{12}\times
C_{11})^a
(C_{12}\times C_{11})^a>,\tag 27
\endalign
$$
with $\ov{C}_{13}^a=C_{13}^a+(C_{12}\times C_1)^a$, and finally
$$
S_2=<C_3^a(\partial_{\mu}A^{\mu
a}-{\alpha\over{2}}C_3^a)-C_2^a\partial_{\mu}\D^\mu
C_1^a+C_2^a\partial_{\mu}C_1^{\mu a} >. \tag 28 $$

It is interesting to note that the latest two terms in $S_1$ (27)
which are of degree four and five in ghost and antighosts fields are due
by the presence of terms of degree two and three in this field in the
gauge fixing function $\chi_{12}^a$ in (23) . This terms were not present in
the gauge fixed action found in [2]. However since they arise from gauge
fixing terms and the remaining term $\D^{\mu}C_{1\mu}^a$ by itself is an
admisible gauge fixing condition the gauge fixed action in [2] is
equivalent to Witten's TQFT. In order to compare with Witten's formulation
we perform the trivial change of variables:
$$
\align
&\psi_{\mu}^a=-iC_{1\mu}^a,\\ &\lambda^a=iC_{11}^a,\\ &\eta^a=-
\ov{C}_{13}^a,\\ &\phi^a=-2iC_{12}^a,\\ &\chi_{\mu\nu}^a=-C_{2\mu\nu}^a, \tag
29
\endalign $$
in $S_1$ (27). The resulting gauge fixed Lagrangean is then exactly $L'$
[1].

We have introduced a new gauge invariant, metric independent, action
which after BRST gauge fixing becomes identical to Witten's TQFT. The
canonical formulation of the action we propose and of the one introduced
by Labastida and Pernici are identical. The algebra of first class
constraints of these actions is the same as the corresponding one for the
theory proposed by Baulieu and Singer and also for the topological BF
theory which certainly describes a non-equivalent topological theory. The
BRST charges, however are different for these three theories.

The gauge action (3) is metric independent allowing a BRST topological
treatment using the BFV [[8,9] off-shell approach. Consequently the BRST
transformation laws we obtain close off-shell, improving the results in
Ref.[2].

The other interesting property we obtain is a new proof that in this
theories the semiclassical approximation is exact. This property was
proven by Witten [1] by explicit use of the particular structure of the
action, being an anticonmutator involving the BRST charge. Instead we
use the structure of the BRST charge (17), which is linear in the conjugate
momenta. From this property it can be shown by inspection that the
terms in Eq.(20) and Eq.(21) which do not involve the gauge fixing
condition are linear in the conjugate momenta. This implies that the gauge
coupling constant ${1\over{e^2}}$ of the effective action can be absorbed
into the momenta by a change of variables in the functional integral. We
are then left with the same effective action but in terms of a new gauge
fixing condition. Since the functional integral is independent of the gauge
fixing condition it implies the independence of any correlation
function on the coupling constant. We may then take the limit $e
\to 0$ as in Witten's argument and evaluate around the classical
minima. This proof may be extended to all BF theories in any dimension
[11].

\vskip 3cm
\noi
{\bf REFERENCES}
\vskip .3cm

\item{[1]}E. Witten, {\it Commun. Math. Phys.} {\bf 117} (1988) 353.
\item{[2]}J. M. F. Labastida and M. Pernici, {\it Phys. Lett.} {\bf B212}
(1988) 56.
\item{[3]}L. Baulieu and I. M. Singer, {\it Nucl. Phys.} (Proc.
Suppl.) {\bf 5B} (1988) 12.
\item{}Y. Igarashi, H. Imai, S. Kitakado and H. So, {\it Phys. Lett.}
{\bf B227} (1989) 239.
\item{}C. Arag\~{a}o and L. Baulieu, {\it Phys. Lett.}
{\bf B275} (1992) 315.
\item{[4]}A. S. Schwarz, {\it Commun. Math.
Phys.}{\bf 67} (1979) 1.
\item{[5]}D. Birmingham, M.Blau, M. Rakowski and
G. Thompson, {\it Phys. Rep.}{\bf209} (1991) 129.
\item{[6]}I. A. Batalin and G. Vilkovisky, {\it Phys. Rev.} {\bf D28}
(1983) 2567.
\item{[7]}G. T. Horowitz, {\it Commun. Math.Phys.}{\bf125} (1989) 417.
\item{[8]}I. A. Batalin and E. Fradkin, {\it Phys. Lett.} {\bf B122} (1983)
157;{\it Phys. Lett.} {\bf 128} (1983) 307.
\item{[9]}M. I. Caicedo and A. Restuccia, {\it Class. Quan. Grav} {\bf 10 }
(1993) 833; {\it Phys. Lett.} {\bf B307 } (1993) 77.
\item{[10]}M. Blau and G. Thompson, {\it Ann.Phys.} {\bf 205}
(1991) 130.
\item{[11]}M. Caicedo, R. Gianvittorio, A. Restuccia and J. Stephany, in
preparation.

\end